\documentclass[aps,prb,reprint,groupedaddress]{revtex4-1}

\usepackage{graphicx}
\usepackage{amsmath}
\usepackage{color}

\begin{document}

\title{Fracturing graphene by chlorination: a theoretical viewpoint}

\author{M. Ij{\"{a}}s}
\email{mari.ijas@aalto.fi}
\author{P. Havu}
\author{A. Harju}
\affiliation{Department of Applied Physics and Helsinki Institute of Physics, Aalto University School of Science, Espoo, Finland}

\date{\today}


\begin{abstract}
Motivated by the recent photochlorination experiment [B. Li~\emph{et al.},~ACS~Nano~5,~5957~(2011)], we study theoretically the interaction of chlorine with graphene.  In previous theoretical studies,  covalent binding between chlorine and carbon atoms has been elusive upon adsorption to the graphene basal plane.  Interestingly, in their recent experiment,  Li~\emph{et al.} interpreted their data in terms of chemical bonding of chlorine on top of the graphene plane, associated with a change from sp$^2$ to sp$^3$ in carbon hybridization and formation of graphene nanodomains.  We study the hypothesis that these domains are actually fractured graphene with chlorinated edges, and compare the energetics of chlorine-containing graphene edge terminations, both in zigzag and armchair directions,  to chlorine adsorption onto infinite graphene. Our results indicate that edge chlorination is favored over adsorption in the experimental conditions with radical atomic chlorine and that edge chlorination with sp$^3$-hybridized edge carbons is stable also in ambient conditions. An \emph{ab initio} thermodynamical analysis shows that the presence of chlorine is able to break the pristine graphene layer. Finally, we discuss the possible effects of the silicon dioxide substrate on the chlorination of graphene.
\end{abstract}


\maketitle

\section{Introduction}
Graphene, the two-dimensional allotrope of carbon, has been a topic of intense study ever since its experimental discovery in 2004. As the properties of pristine graphene are becoming increasingly well understood, the research community has turned its attention toward the modification of graphene properties. Graphene is a zero band gap material that has a linear dispersion in the chiral electronic bands near the Fermi level, which causes some of its extraordinary properties such as high electron mobility. The lack of a band gap in bulk graphene poses, however, limitations for its use in some applications, such as in graphene-based transistors.\cite{Avouris}  Cutting graphene into narrow ribbons introduces a band gap.\cite{Han} An alternative way of introducing a gap is chemical modification that occurs either at graphene edges, or through adsorption or chemical binding to the carbon plane. For instance, through hydrogen\cite{Sofo,  Elias, Ryu} or fluorine\cite{Robinson, Zboril, Nair} attachment onto the basal plane of graphene a band gap can be induced. Furthermore, the properties of graphene nanoribbons can be modified through edge-terminating groups \cite{Li-Zhou-Shen-Chen, Peng-Velasquez,   Rosenkranz, Seitsonen, Wagner} and even half-metallicity can enter through inequivalent edges.\cite{Li-Huang-Duan, Wu}

This far, only the lightest one of the halogen atoms, fluorine, has been extensively studied in the context of interaction with graphene, both theoretically\cite{Maruyama-Kusakabe, Bhattacharya, Leenaerts, Klintenberg,  Medeiros, Wehling, Lee-Cohen-Louie, Rudenko}  and experimentally.\cite{Robinson, Zboril, Nair} The structures of fully fluorinated graphene and hydrogenated graphene are similar; in both the covalently bound heteroatoms changing the carbon hybridization from $sp^2$ to $sp^3$  and strongly corrugating the graphene layer.\cite{Sofo, Elias, Ryu} Instead, calculations with heavier halogens (Cl, Br, and I) on graphene with all carbon atoms halogenated have predicted ionic binding to graphene with no change in the carbon hybridization, retaining the planarity of the carbon network, and no induced band gap.\cite{Klintenberg, Medeiros} The carbon-halogen distances for these structures are around 4~\AA{}, much higher than typical covalent bond lengths. Additionally, studies on adsorption of both isolated heavier halogen atoms \cite{Zhou, Wehling} and halogen molecules \cite{Rudenko} on graphene have predicted no covalent bonding between the halogen atoms and graphene carbons. For graphene nanoribbons (GNRs), some structures with halogen-carbon binding at the edges of a zigzag GNR (ZGNR)  at low edge coverage have been reported.\cite{Lee-Cohen-Louie} 
 
Recently, Li \emph{et al.}\cite{Li} studied graphene-chlorine interactions by using radiation to break Cl$_2$ molecules to reactive chlorine radicals. Based on XPS and Raman spectra, they report graphene basal plane chlorination associated with sp$^3$ hybridized carbon atoms, the formation of graphene nanodomains that are approximately 30-50~nm in lateral dimension, the appearance of a band gap greater than 45~meV, and estimate the degree of chlorination to 8~atom-\%. In another recent experimental report, Wu~\emph{et al.}\cite{Wu-Xie-Li} report two different reaction regimes for graphene exposed to Cl$_2$ plasma. For short exposure times, the conductivity of graphene is increased related to p-doping, only a weak defect-related Raman peak is seen, and pristine graphene is recovered upon annealing. For long exposure, an irreversible change in the graphene properties is reported, associated with formation of chlorine binding defect patches as seen in transmission electron microscopy images. {Thus, in both of these experiments, changes in the structure of graphene are observed upon chlorination.}

\begin{table*}
 \caption{\label{table:Cl_ads} Chlorine binding energies [Eq.~(\ref{eq:Eb_Cl}), in eV per Cl atom] and chlorine equilibrium distance from the carbon plane  for different supercell sizes and initial positions of the chlorine atom. The calculations starting from the hollow position that converge to top positions are excluded from the table.}
 \begin{tabular}{lccccccccc}
\hline
\hline
      & bridge   &   &    & hollow&  &     & top&  &    \\
      & $E_{B,\mathrm{Cl}}$  & $E_{B,\mathrm{Cl_2}}$ & $d$& $E_{B,\mathrm{Cl}}$  & $E_{B,\mathrm{Cl_2}}$ & $d$  &$E_{B,\mathrm{Cl}}$  & $E_{B,\mathrm{Cl_2}}$& $d$  \\
    &(eV/Cl) & (eV/Cl) & (\AA{}) & (eV/Cl) & (eV/Cl) & (\AA{})& (eV/Cl) & (eV/Cl) & (\AA{})\\
\hline
4$\times$4  &-1.040 &0.703&2.93 & -     & -   & -    &-1.049 &0.695 &2.86 \\
6$\times$6  &-1.159 &0.584&3.10 &-1.142 &0.601&3.23 &-1.161 &0.582 &3.05 \\
8$\times$8 &-1.210 &0.533&3.12 &-1.196 &0.547&3.22 &-1.212 &0.532 &3.08 \\
12$\times$12&-1.244 &0.499&3.17 & -      & -    &  -    &-1.248 &0.496 &3.12 \\
\hline
\hline
\end{tabular}
\end{table*}

The differences in these experiments may be due to different chlorination procedure and  graphene quality. The purpose of this study is to address this discrepancy between  the binding of Cl on graphene in the photochlorination experiment\cite{Li} and previous theoretical calculations,\cite{Medeiros, Klintenberg,Zhou, Wehling} commenting also the very recent experimental observations by Wu~\emph{et al.} in Ref.~\onlinecite{Wu-Xie-Li}. We propose that the presence of atomic chlorine can lead to the formation of edges in graphene, so that the nanodomains observed by Li~\emph{et al.}\cite{Li} could actually be chlorine-edged nanostructures, and compare the chlorinated edge formation to adsorption onto bulk graphene. By considering a large number of chlorine-containing edge terminations in our density-functional theory calculations,  we show that chlorine atoms preferably bind to the graphene edges, rather than onto the basal plane of graphene. Using an \emph{ab initio} thermodynamics approach, we find that only chlorinated armchair (AC) edges are stable with respect to pristine graphene, and that in the presence of radical chlorine, graphene may spontaneously break into structures with chlorinated armchair edges.  Additionally, we discuss the possible effects that a SiO$_2$ substrate can have on chlorination.

\section{Computational details}

Our calculations were performed using the density-functional theory (DFT) with a van der Waals (vdW) correction,\cite{Tkatchenko} implemented in the all-electron density-functional code, developed at the Fritz-Haber Institute (FHI), called FHI-aims.\cite{AIMS} Double numeric plus polarization basis set of numerical atom-centered orbital basis functions and the PBE \cite{PBE} exchange correlation functionals were used.  The structural relaxation (with the vdW correction) was converged until the forces acting on the atoms were less than 0.001~eV/\AA{} and the electronic degrees of freedom were converged below $10^{-6}$~eV. Unless otherwise mentioned, no spin polarization was taken into account, as in previous calculations chlorine adsorption has not been shown to induce spin polarization.\cite{Zhou}

The optimized lattice constant, $1.42$~\AA{}, of an infinite graphene sheet was used in all calculations in the absence of a substrate. In order to avoid interaction between the periodic images, a vacuum layer of 15~\AA{} was placed between the graphene layers. The Brillouin zone sampling was adjusted to the supercell size, corresponding to a 48$\times$48 k-point mesh for the graphene primitive unit cell.  In the calculation of binding energies, the reference state for carbon was chosen as the infinite pristine graphene sheet, and for hydrogen a hydrogen molecule in the gas phase. For chlorine, both molecular and atomic reference states were used, as in the photoclorination experiment radiation is used to break the clorine molecules into atomic chlorine radicals.\cite{Li} Thus, the formation and binding energies calculated with respect to atomic chlorine corresponds to reaction conditions, and the molecular to ambient conditions.

\section{Results}
\subsection{Chlorine adsorption on infinite graphene \label{sec:Clads}}

First, we study the binding of one chlorine atom on the infinite graphene sheet. In previous calculations, no covalent binding between graphene carbon and chlorine atoms that could change the hybridization of carbon atoms from sp$^3$ to sp$^2$ has been found. Instead, the adsorption was of ionic type, associated with charge transfer between graphene and the adsorbate, that has the lowest-energy adsorption positions  on top of a carbon atom, \cite{Wehling} or on the bridge site between two carbon atoms. \cite{Wehling, Zhou} The carbon-chlorine distances were around 3~\AA{},\cite{Zhou} clearly longer than the covalent bond length between carbon and clorine, 1.72-1.85~\AA{}.\cite{Allen} The adsorption energies in these calculations were $-0.3$~eV/Cl \cite{Zhou} or $-0.80$~eV/Cl,\cite{Wehling} the chlorine reference state being atomic chlorine. 

In our calculations, we consider three initial positions for chlorine atoms: on top of a carbon atom~(t), bridge between two carbon  atoms~(b) and in the center of a carbon hexagon~(h). As periodic supercells are used in the calculation, we address the finite size effects by calculating the binding energy for the single Cl atom absorption to graphene for different supercell sizes. The binding energy is given by
\begin{equation} \label{eq:Eb_Cl} E_{B,\mathrm{Cl}} = E_{\mathrm{Cl-gra}}-E_{\mathrm{gra}}-E_{\mathrm{Cl}}, \end{equation}
where $E_{\mathrm{Cl-gra}}$ is the total energy of the graphene supercell with a single chlorine atom,  $E_{\mathrm{gra}}$ the energy of pristine graphene in a supercell of
the same size, and $E_{\mathrm{Cl}}$ the energy of a chlorine atom. When molecular reference for chlorine is used, $E_{\mathrm{Cl}}$ is substituted by $E_{\mathrm{Cl}_2}/2$.

Table~\ref{table:Cl_ads} shows our results for the atomic chlorine binding energy and the equilibrium distance for different supercell sizes, ranging from a 4$\times$4 supercell (32 carbon atoms) also used in Ref.~\onlinecite{Wehling}, to a 12$\times$12 supercell (288 carbon atoms). Our results qualitatively agree with those of Ref.~\onlinecite{Wehling}, giving little difference in adsorption energies between top and bridge adsorption sites and no covalent binding or noticeable distortion of the graphene plane. Finite-size effects are clearly significant both for the binding energies and for the chlorine equilibrium distance but as the energies are converging, we use the values calculated using the 12$\times$12 supercell for comparisons. As seen from the binding energies (Table~\ref{table:Cl_ads}), chlorine physisorption on all sites is energetically feasible in radical chlorine conditions ($E_{B,\mathrm{Cl}}$), whereas adsorption does not occur when recombination to molecular chlorine is possible ($E_{B,\mathrm{Cl}_2}$),  and metastable adsorbed structures are prone to Cl$_2$ desorption. The lower values for the adsorption energy ($-1.25$~eV for top site adsorption calculated for the 12$\times$12 supercell) in comparison to Ref.~\onlinecite{Wehling} are due to the inclusion of the additional adhesion-providing van der Waals correction. Using the same supercell size as in Ref.~\onlinecite{Wehling} and excluding the van der Waals correction, we get the chlorine adsorption energy $E_{B,\mathrm{Cl}} = -0.76$~eV/Cl, agreeing with the value $-0.80$~eV/Cl that Wehling~\emph{et al.}\cite{Wehling} reported. 

\begin{table}
\caption{\label{table:subst} The chlorination of graphene on SiO$_2$ for different substrate terminations in the 2$\times$2 and 4$\times$4 graphene supercells. The C-Cl bond distance for adsorbed chlorine $(d_{C-Cl})$ and the chlorine binding energy with respect to radical $(E_{B, \mathrm{Cl}})$ and molecular chlorine $(E_{B, \mathrm{Cl}_2})$ are given for all surface terminations.}
 \begin{tabular}{lccc}
\hline
\hline
termination  & $E_{B, \mathrm{Cl}}$ & $E_{B, \mathrm{Cl}_2}$&	 $d_{C-Cl}$\\ 
		   &  (eV/Cl)		  & (eV/Cl)		    & (\AA{})		  \\ 	
\hline
O$_{2\times2}$     & -0.920 &0.530 &1.83 \\
O$_{4\times4}$     &0.063 &1.513&1.83\\
& & & \\
Si$_{2\times2}$      &-0.901 &0.549&1.80	\\
Si$_{4\times4}$      &-0.782 &0.668 &3.08	\\
& & & \\
OH$_{2\times2}$       &-1.622  &-0.173&1.90	\\  
OH$_{4\times4}$       &-0.827 &0.623&1.90	\\  
& & & \\
rec-O$_{2\times2}$  &-0.432&1.018&2.93 	\\
rec-O$_{4\times4}$  &-0.823 &0.625 &3.02 	\\
\hline  
\hline
 \end{tabular}
\end{table}

Wu~\emph{et al.}\cite{Wu-Xie-Li} studied one-sided graphene chlorination also theoretically by calculating the chlorine binding energy on pristine graphene at different coverages. Using atomic chlorine as the reference state, they find that binding is energetically stable, indicated by negative binding energies.  We test this by calculating the binding energy for the structure corresponding to their calculation, using both atomic and molecular reference for chlorine. For 1/4~monolayer coverage, our result for the binding energy with respect to atomic chlorine is with (without) vdW correction $-0.44$~($-0.29$)~eV/Cl, agreeing well with approximately $-0.25$~eV reported by Wu~\emph{et al.} in Fig.~4(a) of Ref.~\onlinecite{Wu-Xie-Li}. Using a molecular reference state, however, the binding energies increase to 1.01~(1.16)~eV/Cl and the adsorbed structure is unstable, and higher in energy than our result for the adsorption energy of a single chlorine atom, $-1.25$~eV/Cl.

\begin{figure}
   \begin{tabular}{cc}
\begin{tabular}{c} \includegraphics[width=0.45\columnwidth]{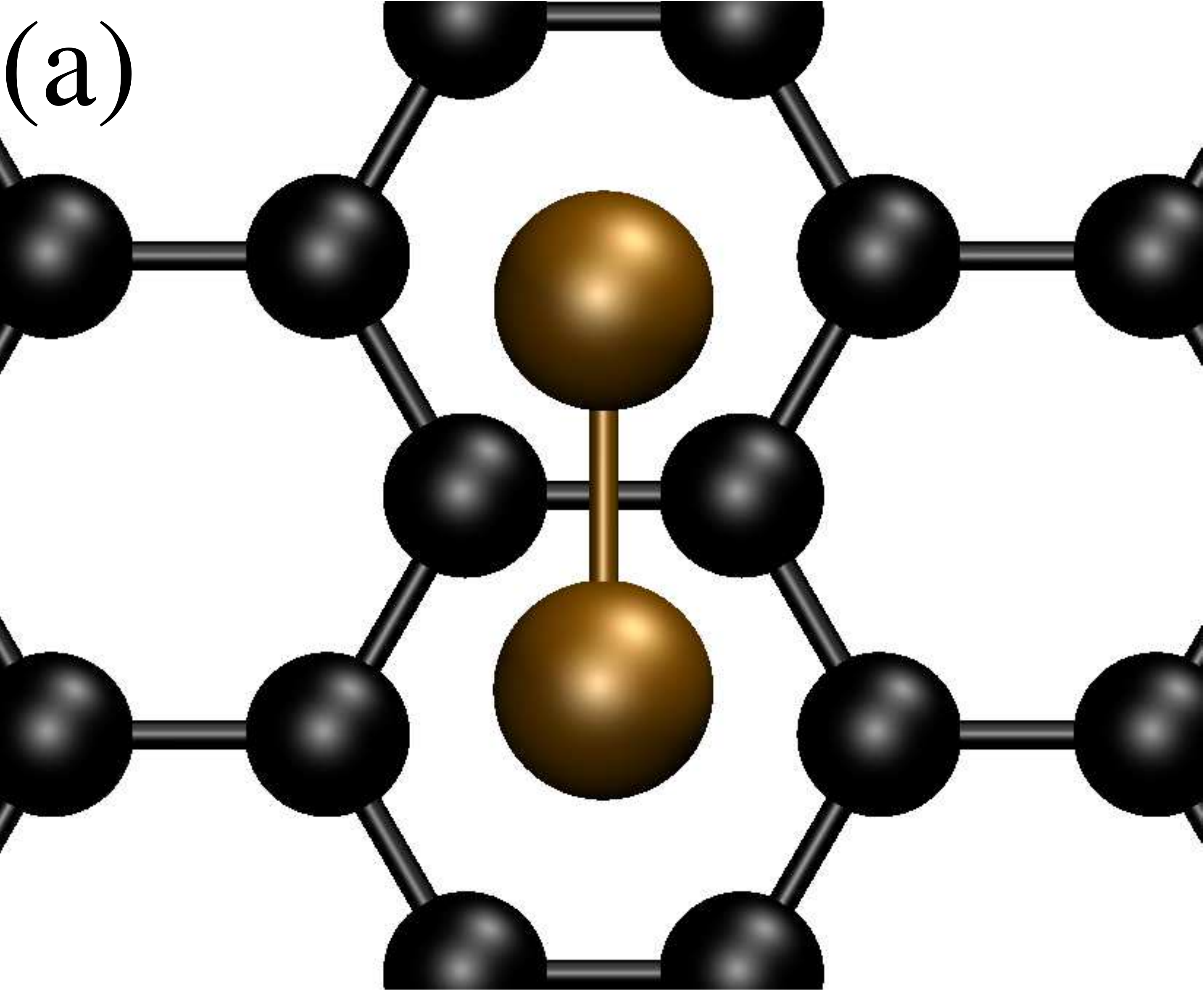}\\ $E_{B, \mathrm{Cl}}$ =-1.867~eV/Cl\\$E_{B,  \mathrm{Cl}_2}$=-0.125~eV/Cl \\ \end{tabular} &
\begin{tabular}{c}  \includegraphics[width=0.48\columnwidth]{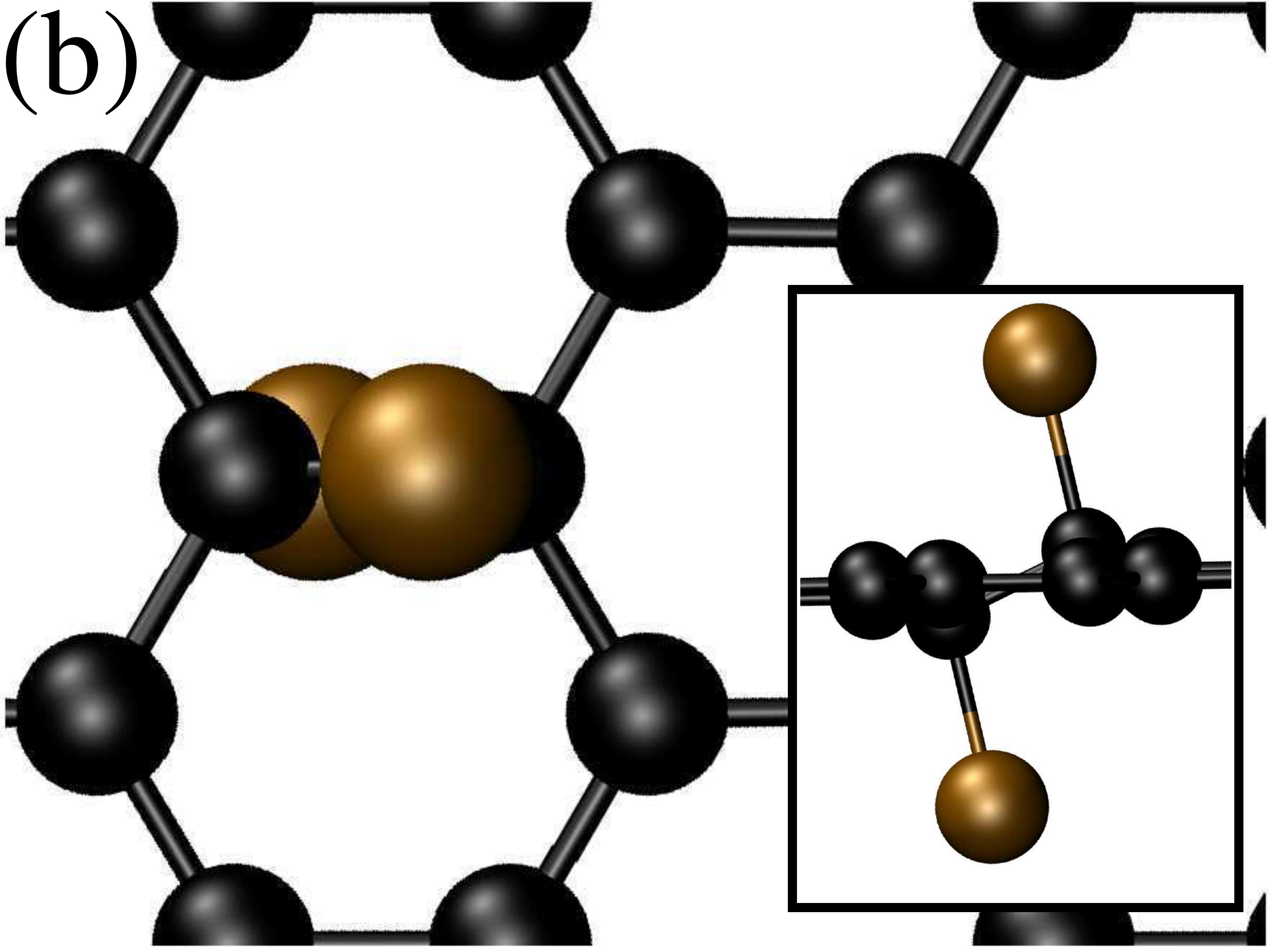}\\$E_{B, \mathrm{Cl}}$ = -1.333~eV/Cl\\$E_{B,  \mathrm{Cl}_2}$=0.410~eV/Cl \\ \end{tabular} \\
\begin{tabular}{c} \includegraphics[width=0.45\columnwidth]{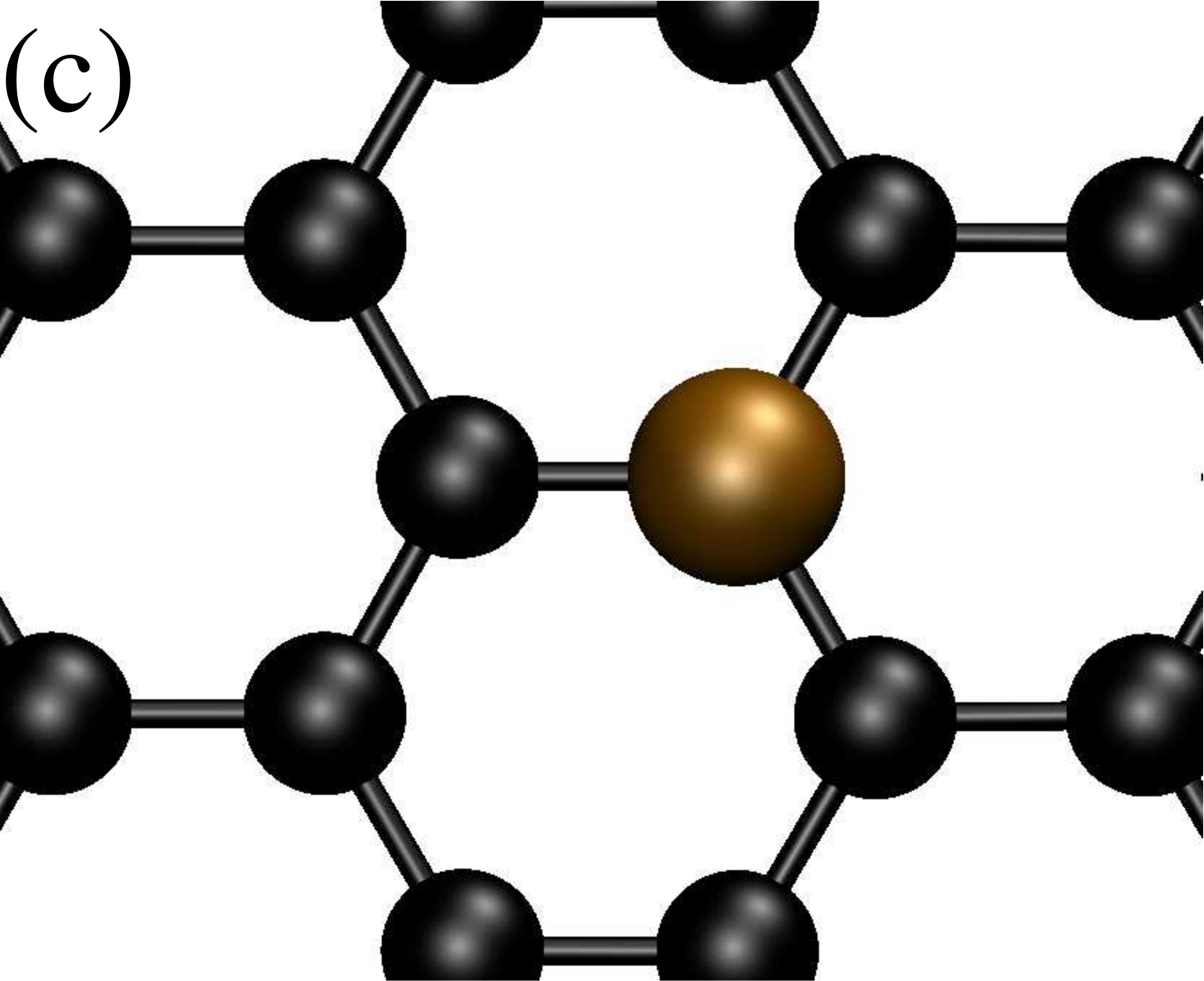} \\ $E_{B, \mathrm{Cl}}$= -1.335~eV/Cl\\ $E_{B,  \mathrm{Cl}_2}$= 0.408~eV/Cl\\ \end{tabular} &
\begin{tabular}{c}  \includegraphics[width = 0.48\columnwidth]{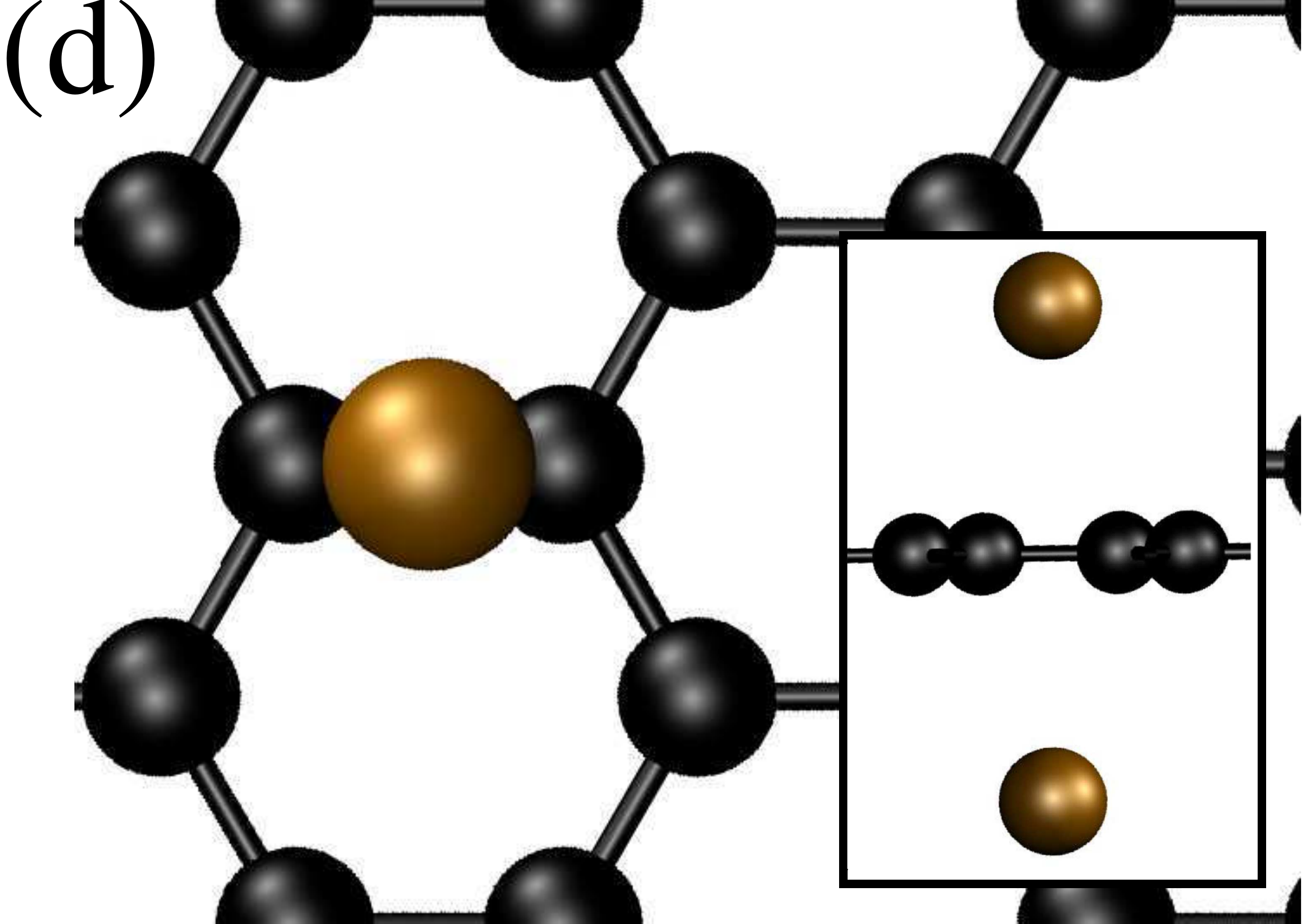} \\$E_{B, \mathrm{Cl}}$= -1.277~eV/Cl\\$E_{B,  \mathrm{Cl}_2}$=0.466~eV/Cl\\ \end{tabular}\\ 
 \end{tabular}
 \caption{\label{fig:Cl2gra} (Color online) The geometries for the different two-chlorine configurations. (a)~Cl$_2$ molecule (b)~on  the opposite sides of the carbon layer, bridge site, bound structure [inset: side view] (c)~opposite sides of the carbon layer, top site (d)~on  the opposite sides of the carbon layer, bridge site, adsorbed structure [inset: side view]. Below the structures, the binding energy in radical and ambient conditions~($E_{B, \mathrm{Cl}}$ above $E_{B,  \mathrm{Cl}_2}$) is given per chlorine atom. Atom colors: Brown/gray--chlorine, black--carbon.}
\end{figure}

More complicated adsorption structures are, of course, possible, as well as molecular adsorption that has previously been studied in more detail by Rudenko~\emph{et   al.} in Ref.~\onlinecite{Rudenko}. We restrict ourselves to a pair of Cl atoms placed in the 8x8 graphene supercell.  We note that if the distance between two chlorine atoms is initially less than  $\sqrt{7}$ times the carbon-carbon distance in graphene, they attract each other and spontaneously form chlorine molecules during the relaxation. Our results for the adsorption energy with respect to molecular chlorine and equilibrium distance {of} a Cl$_2$ molecule agree well with the results of Rudenko~\emph{et   al.},\cite{Rudenko} giving an equilibrium distance of 3.49~\AA{} and binding energy $-0.250$~eV/Cl$_2$, compared to 3.58~\AA{} and $-0.259$~eV/Cl$_2$ by Rudenko~\emph{et al.} in Ref.~\onlinecite{Rudenko}. The final adsorption position for a chlorine molecule is shown in Fig.~\ref{fig:Cl2gra}(a). Instead, for initial configurations where the distance between the chlorine atoms is longer than  $\sqrt{7}$ times the carbon-carbon distance, the mutual repulsion between the Cl atoms leads to maximum interatomic distance between the chlorine atoms within the finite supercell.  

In the photochlorination experiment by Li~\emph{et al.}~\cite{Li}, graphene and chlorographene were deposited on the SiO$_2$ substrate. In the case of graphene hydrogenation, the substrate may affect the stability of adsorbed structures, as recently reported for hydrogenated graphene, graphane, on SiO$_2$.\cite{Havu} In order to take this possibility into account, we calculated the chlorine binding energies on the $\alpha$-phase SiO$_2$ surface.  As SiO$_2$ is amorphous, we considered  four different surface terminations: oxygen-terminated (O), silicon-terminated (Si), hydroxyl-terminated (OH) and reconstructed oxygen terminated (rec-O) surface, for which the initial geometry was taken from Ref.~\onlinecite{Goumans}. The lattice mismatch between the optimized substrate and graphene was only 1.3\% when a 2$\times$2 supercell formed from graphene primitive unit cells with the optimized lattice constant corresponding to freestanding graphene was used. Three SiO$_2$ unit cells were used in the slab, corresponding to a layer of 15.8~\AA{}, that was either symmetrically chlorinated from both sides, or terminated with an ideal graphene-substrate interface. The chlorine atom, graphene, and the three uppermost atom layers of the substrate  were relaxed until forces acting on atoms were less than 0.01 eV/\AA{}.  The convergence criterion for the electronic degrees of freedom was 10$^{-6}$~eV and a k-point mesh of 6$\times$6$\times$1 was used to sample the Brillouin zone.

For chlorine adsorption above the graphene plane, we find that in the radical chlorine environment, binding to the surface is energetically feasible on all surface terminations for the 1/8 monolayer chlorine coverage (2$\times$2 graphene supercell). In contrast to adsorption onto freestanding graphene, we find covalent bonding between the chlorine atom and graphene on all surfaces apart from the rec-O surface, as seen in the carbon-chlorine bond lengths shown in Table~\ref{table:subst}.   The absorption onto the OH surface is energetically most favored, as in the case of hydrogen adsorption.\cite{Havu} In the ambient environment, using molecular chlorine as the reference state, the adsorption energy per chlorine atom is negative only for the OH-terminated surface, the adsorption energy being $-0.17$~eV/Cl. This, however, requires an ideally flat, crystalline SiO$_2$ surface and the effect of surface roughness is beyond the scope of this article.

At lower coverages, modeled by increasing the supercell size to correspond to 4$\times$4 graphene unit cells (and 2$\times$2 SiO$_2$ unit cells), scaling the k-point mesh correspondingly, we notice that on those surfaces (O and OH) that form bonds with chlorinated graphene,  the absorption energies increase  as the distance between the adsorbates increases (Table~\ref{table:subst}) and adsorption becomes less favorable.  On {the} rec-O {surface}, on the contrary, the absorption energy decreases.  On the Si surface,  graphene binds both to the substrate and to the chlorine atom in the 2$\times$2 supercell but increasing the supercell to 4$\times$4 leads to change in the binding to physisorption, seen in Table~\ref{table:subst} as increase in the C-Cl distance from 1.80~\AA{} to 3.08~\AA{}. As chlorine binds to graphene on the O and OH surfaces, the planarity of the carbon network is disturbed and binding to the substrate provides additional stabilization. In this case, it is beneficial to have adsorbates relatively close to each other, keeping the carbon backbone similar to the chair configuration of graphane. The Si surface is more inert and unable to provide this stabilization at low coverage. On the other hand, graphene on the rec-O surface without chemical bonding between substrate atoms and the carbons resembles a freestanding graphene membrane with adsorbates. Like for the freestanding graphene, there is a  repulsion between the chlorine atoms in neighboring periodic images,\cite{repulsion} and increasing the distance between the adsorbates stabilizes the structure. At this lower coverage, chlorinated graphene is no longer stable on any of the surfaces and the discrepancy between the experiment and theoretical calculations can not be simply explained by substrate effects. We exclude the study of chlorine adsorption on defected graphene from this study.

It could, in principle, be possible that some chlorine intercalates between the graphene layer and the substrate through cracks and boundaries in the graphene layer, thus enabling chlorine attachment to graphene from both sides of the carbon network.  Returning to graphene in the absence of a substrate, we probe this by placing two chlorine atoms on the opposite sides of the graphene layer, either on top of a single carbon atom, on neighbouring carbon atoms or symmetrically on both sides of a bridge site. Chlorine atoms initially placed on the top of neighboring carbons on both sides of the graphene layer relax to a bridge position. The corresponding {two-chlorine} geometries are shown in Figs.~\ref{fig:Cl2gra}(b), \ref{fig:Cl2gra}(c), and \ref{fig:Cl2gra}(d) that also give the binding energy per chlorine atom in radical and ambient conditions~($E_{B, \mathrm{Cl}}$ above $E_{B,  \mathrm{Cl}_2}$, respectively). The lowest-energy configuration corresponds to double-sided top position [\ref{fig:Cl2gra}(c)], with a shorter equilibrium distance than for a single adsorbed chlorine atom ($\approx$ 2.5~\AA{} compared to $\approx$ 3.1~\AA{}). The double-sided bridge adsorption position with a slight distortion of the graphene layer [Fig.~\ref{fig:Cl2gra}(b)] is very close in energy, the difference in $E_{B, \mathrm{Cl}}$ being only few meV. Instead, bridge adsorption that retains graphene planarity [Fig.~\ref{fig:Cl2gra}(d)] lies approximately 0.05~eV higher in energy. 

The binding energies per chlorine atom for the Cl$_2$ molecules in ambient conditions (Fig.~\ref{fig:Cl2gra} below the structures), indicate that apart from molecular adsorption, in ambient conditions chlorine binding in pairs on pristine graphene is not feasible and should be prone to desorption processes. Being unable to explain the experimental observations by chlorine adsorption and binding onto the graphene basal plane with bond formation of chlorine with sp$^3$-hybridized carbon seen {in the experiment}, we thus turn our attention to the binding of chlorine to graphene edges.

\begin{figure*}
\includegraphics[width = 0.55\textwidth, angle=90]{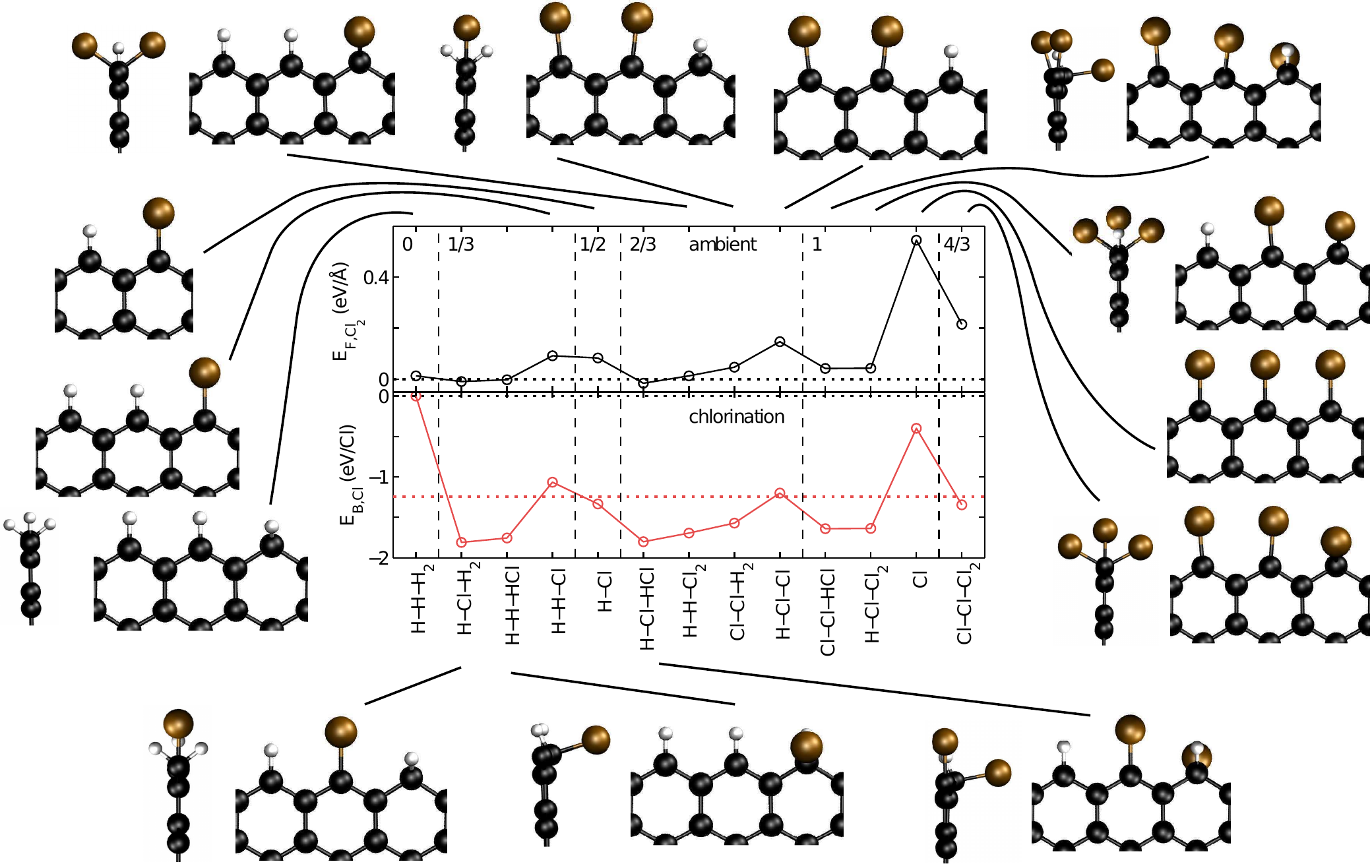}
\caption{\label{fig:zgnr8} (Color online) The edge formation energy $E_{F, \mathrm{Cl}_2}$ in ambient conditions (upper panel) and binding energy per chlorine atom  $E_{B, \mathrm{Cl}}$ (lower panel) in radical chlorine environment for the different zigzag terminations. In the upper panel, the dotted horizontal line shows stability with respect to pristine graphene ($E_{\mathrm{B,Cl_2}}=0$), and in the lower panel, the red dotted line shows the adsorption energy of a single chlorine atom on pristine graphene in the absence of substrate (12$\times$12 supercell, $E_{F, \mathrm{Cl}}=-1.25$~eV/Cl). The geometrical structures are connected to the graph from below (above) if they are (un)stable in ambient conditions. Vertical dashed lines separate groups of different degree of chlorination, expressed as chlorine atoms per edge carbon. Brown/gray: chlorine, black: carbon, white: hydrogen.}	
\end{figure*}

\subsection{Cl in zigzag graphene edges}

As a model for zigzag-terminated (ZZ) graphene edges, we consider an eight zigzag rows wide ZGNR (8-ZGNR, width $\approx$15.6~\AA{}). For fully hydrogen-terminated ribbons, this corresponds to well converged edge energy with respect to the ribbon width.\cite{Wassmann-JACS} We checked this also for some of the chlorinated structures, finding that in the width range 15--28~\AA{}, the formation energies remain essentially constant.  Approximately 20~\AA{} of vacuum was placed between adjacent ribbons to avoid interaction due to periodic boundary conditions.  All atoms in the ribbons were relaxed and the Brillouin zone was sampled only in the direction along the ribbon axis. Comparisons with
systems in which the carbon atoms in the middle of the ribbons were fixed and only carbon atoms belonging to the edge hexagons were relaxed showed only energy differences of order 10~meV in total energies.

\begin{table}[!t]
 \caption{ \label{table:8-zgnr} The energetics for different zigzag edge terminations: the formation energy per unit length of the edge ($E_F$) and the binding energy per chlorine atom~($E_{B}$), both with radical chlorine~($_{,\mathrm{Cl}}$) and with molecular chlorine~($_{,\mathrm{Cl_2}}$) as the reference state. Also the C-Cl bond lengths  are reported.  The ordering is based on edge stability with respect to molecular chlorine~($E_{F,\mathrm{Cl}_2}$).}
\begin{tabular}{lccccc}
 \hline
 \hline
system & $E_{F,\mathrm{Cl}}$ &$E_{B,\mathrm{Cl}}$& $E_{F,\mathrm{Cl_2}}$ &$E_{B,\mathrm{Cl_2}}$&$d_{\mathrm{C-Cl}}$\\
  & \footnotesize{(eV/\AA{})} &\footnotesize{(eV/Cl)} &  \footnotesize{(eV/\AA{})} &\footnotesize{(eV/Cl)} &\footnotesize{(\AA{})}\\
\hline
\footnotesize{H-Cl-HCl}	&-0.487&-1.802&-0.016 &-0.059 &1.88, 1.74 \\
\footnotesize{H-Cl-H$_2$}	&-0.245&-1.810&-0.009 &-0.066 &1.74 \\
\footnotesize{H-H-HCl} 	&-0.237&-1.755&-0.002 &-0.012 &1.88         \\
\footnotesize{H-H-Cl$_2$} 	&-0.458&-1.695& 0.013 & 0.049 &1.83   \\
\footnotesize{H-H-H$_2$}	& 0.013& -    & 0.013 & -     & -       \\
\footnotesize{Cl-Cl-HCl}	&-0.665&-1.640& 0.042 & 0.103 &1.74,1.73,1.88\\
\footnotesize{H-Cl-Cl$_2$} 	&-0.664&-1.638& 0.043 & 0.105 &1.73, 1.83  \\
\footnotesize{Cl-Cl-H$_2$} 	&-0.425&-1.571& 0.047 & 0.173 &1.74      \\
\footnotesize{H-Cl} 		&-0.271&-1.335& 0.083 & 0.408 &1.73      \\
\footnotesize{H-H-Cl} 		&-0.144&-1.066& 0.092 & 0.678 &1.73      \\
\footnotesize{H-Cl-Cl} 	&-0.324&-1.200& 0.147 & 0.544 &1.73         \\
\footnotesize{Cl-Cl-Cl$_2$}	&-0.728&-1.346& 0.215 & 0.397 &1.72,1.86 \\
\footnotesize{Cl} 		&-0.162&-0.399& 0.545 & 1.344 &1.75    \\
\hline
\hline
\end{tabular}
\end{table}

We generate a large number of initial chlorinated configurations by starting both from the monohydrogenated (H-H-H) and the H-H-H$_2$ edge (in which every third edge carbon is dihydrogenated) and by substituting some of the hydrogen atoms by chlorine. As all ribbon atoms are relaxed, some coupling between the two edges of the ribbon may remain. In order to take this into account, we also calculate all different isomers (including stereochemistry) for each edge configuration. The differences in total energy between the isomers are always below 0.1 eV and in many cases they are of order meV, and thus the isomers are reported as a single termination, choosing the lowest-energy isomer as the representative structure for the
termination. The stability of the edges is determined by calculating the edge formation energy per unit length 
\begin{equation} \label{eq:Eform}
E_{F,\mathrm{Cl}} = \frac{E_{\mathrm{Cl-xGNR}} - N_C  E^{\mathrm{C}}_{\mathrm{graphene}}- \frac{N_H}{2} E_{H_2} -   N_{\mathrm{Cl}}E_{\mathrm{Cl}}}{L},  \end{equation}
where $E_{\mathrm{Cl-xGNR}}$ is the energy of the chlorinated ribbon, $N_{\alpha}$ the number of $\alpha$ atoms, $E^{\mathrm{C}}_{\mathrm{graphene}}$ the energy of a single carbon in
pristine graphene, $E_{\alpha}$ the energy of species $\alpha$ and $L$ the length of the ribbon edge. If molecular chlorine is used as the reference state, $N_{\mathrm{Cl}}E_{\mathrm{Cl}}$ is replaced by $N_{\mathrm{Cl}}E_{\mathrm{Cl}_2}/2$. A stable structure with respect to pristine graphene should have a negative edge formation energy. In order to determine whether chlorine prefers to bind to the edge instead of adsorbing onto the basal plane of graphene, we also calculate the chlorine binding energy,
\begin{equation} \label{eq:Ebind} E_{B,\mathrm{Cl}} = \frac{E_{\mathrm{Cl-xGNR}} - N_C   E_{\mathrm{graphene}}- \frac{N_H}{2} E_{H_2} -   N_{\mathrm{Cl}}E_{\mathrm{Cl}}}{N_{\mathrm{Cl}}},  \end{equation} 
and compare it to the  lowest binding energy of a single chlorine atom on an infinite graphene sheet, for which we use the energy corresponding to the top adsorption,   $E_{B, \mathrm{Cl}}=-1.25$~eV.

Table~\ref{table:8-zgnr} shows the edge formation and chlorine binding energies ($E_{F,\mathrm{Cl{/Cl_2}}}$ and $E_{B,\mathrm{Cl{/Cl_2}}}$, respectively) of the chlorinated edges, along with the carbon-chlorine bond length ($d_{\mathrm{C-Cl}}$).  Fig.~\ref{fig:zgnr8} shows $E_{B,\mathrm{Cl}}$ in radical conditions (lower panel) and $E_{F,\mathrm{Cl}_2}$ in ambient conditions (upper panel), along with the geometries of structures of the terminations, depicting some of the data in Table~\ref{table:8-zgnr} pictorially. The structures below the horizontal lines are stable in their own environments (see the caption for details).  Similarly with the lowest-energy hydrogenated zigzag edges,\cite{Wassmann} the lowest-energy structures are commensurate with the H-H-H$_2$ hydrogenation pattern. In all chlorinated edges that modify the H-H-H$_2$ pattern, the binding energy per chlorine atom in radical environment is lower than the adsorption energy of a chlorine atom onto pristine graphene and thus edge binding is favored. Some of the chlorinated H-H-H$_2$ modifications are stable also in ambient conditions (Fig.~\ref{fig:zgnr8} upper panel), although the edge formation energies are very small, of the order $-10^{-3}$~eV/\AA{}. The degree of chlorination of the edge carbon atoms is 2/3 in the lowest-energy configuration (H-Cl-HCl). In  general, there is only little correspondence between the degree of chlorination and edge stability, as seen in Fig.~\ref{fig:zgnr8}.

Looking at the atomic arrangement of the lowest-energy structure, H-Cl-HCl (Fig.~\ref{fig:zgnr8}, bottom right), we notice that one of the chlorine atoms binds to a planar sp$^2$ carbon. The carbon atom binding to both hydrogen and chlorine slightly relaxes upwards towards the chlorine and the chlorine is close to a top adsorption position for the edge carbon.  Previously, Lee~\emph{et al.}\cite{Lee-Cohen-Louie} studied the interaction of monohydrogenated ZGNRs and molecular chlorine and found similar stable structures with some of the edge carbon atoms bound to both chlorine and hydrogen at low chlorine coverage of the edge atoms.  The C-Cl bond lengths give some indication on the nature of binding. Although all C-Cl bond lengths agree fairly well with the average literature value for covalent C-Cl bonds of 1.76~\AA{},\cite{Handbook} two different regimes are to be seen in Table~\ref{table:8-zgnr}. The chlorine atoms are more strongly bound to sp$^2$ hybridized carbon atoms, indicated by the shorter bond lengths 1.72-1.75~\AA{}. The bonds between chlorine atoms and carbon atoms with increased sp$^3$ character are slightly longer, 1.83-1.88~\AA{}. These values are well in line with literature values for carbon-chlorine bond lengths in different chemical environments.\cite{Allen}

Although we use the GNRs only to model graphene edges, we briefly comment on the band gaps of the chlorinated ribbons. ZGNRs with the H-H-H$_2$ hydrogenation pattern are predicted to have a band gap that is also dependent on the relative arrangement of the H$_2$ groups across the ribbon edges,\cite{Wassmann, Wassmann-JACS} whereas the monohydrogenated ribbons are predicted to be metallic in the absence of spin polarization. If spin polarization is taken into account, an antiferromagnetic spin structure between the two graphene sublattices emerges, leading to spin-polarized edge states in the metallic ribbons.\cite{Wassmann-JACS} In the terminations commensurate with the H-H-H$_2$ pattern, the edge carbon atoms {that bind} to two substituents are not a part of the graphene $\pi$ electron network and the spin-polarized edge state is destroyed. The substitution of edge hydrogens by chlorine atoms does not change this picture. The band structures in the absence of spin polarization are metallic if the edge carbons are monoterminated, and in the case of a H-H-H$_2$ modification a band gap is {found}. The presence of chlorine does not seem to disturb the edge state and modifies only slightly the band gap values. For instance, the band gap of a 8-ZGNR  with H-H-H$_2$ termination is 0.92~eV whereas for the H-Cl-H$_2$ termination the band gap is 0.91~eV. 

As the realisticity of spin-polarized edge states in ZGNRs has been debated\cite{Kunstmann}, we note only briefly that the chlorinated, monosubstituted ZGNRs show spin polarized edge states that are slightly lower in energy than the nonmagnetic ones. The spin moments on chlorine-binding carbon atoms are in most cases slightly enhanced in comparison to the hydrogen-bound ones. As an extreme example, in the H-Cl-Cl terminated ribbon the spin magnitude at the   edge carbons are 0.16~$\mu_B$ and 0.18~$\mu_B$ for the carbon atoms bound  to hydrogen and chlorine, respectively. An exception is the fully chlorinated edge (named Cl), where the maximal spin moment is only  0.06~$\mu_B$ and the spin polarization is thus strongly suppressed  in comparison to the monohydrogenated edge with maximal spin moments   of 0.20~$\mu_B$. This may be due to significant chlorine  contribution to the density of states and an extra dispersive energy  band near the Fermi level in this structure (not shown).

\subsection{Cl in armchair graphene edges}

It is surprising that even though the modification of ZGNRs with different edge-terminating functional groups and the resulting effect on their electronic and magnetic properties has been a field of intense study (see, eg., Refs.~\onlinecite{Wu},~\onlinecite{Park}, and~\onlinecite{Ramasubramaniam}), the edge functionalization of armchair graphene nanoribbons (AGNRs) has caught the attention of the scientific community only recently.\cite{Peng-Velasquez, Seitsonen, Wagner, Rosenkranz} Before that, AGNR ribbons have been assumed to be fully planar. Rippling of the AGNR edge in the presence of non-hydrogen terminating groups has, however, been predicted in recent theoretical works\cite{Rosenkranz,  Wagner} and we take this possibility  into account by using a doubled supercell along the ribbon axis in our calculations. In principle, ripples with a longer wavelength could also be possible, but our choice will most likely capture the essential physics of the system regarding the interaction between the edges and chlorine. Indeed, for all singly-terminated ribbons apart from the plain hydrogen termination, we found structures with edge rippling to be lower in energy than fully planar structures with the same termination. 

Previously, the interaction of chlorine with armchair edges has been only briefly studied. Lee~\emph{et al.}~\cite{Lee-Cohen-Louie} studied the interaction of chlorine molecules with GNR edges but found only weak adsorption of chlorine molecules near the ribbon edge with no covalent binding. Wagner~\emph{et al.}~\cite{Wagner} report rippling of the monochlorinated edge (the Cl edge in our nomenclature) but they do not comment on its energetic stability.

In order to study chlorinated structures of an armchair-terminated graphene edge, we use a 13-AGNR (width $\approx$ 14.8~\AA{}) as the model system. The electronic and energetic properties of AGNRs are known to oscillate with respect to the ribbon width.\cite{Barone,Wassmann-JACS} Like in the case of ZGNRs, we consider a large number of chlorinated edge structures created both from the monohydrogenated and dihydrogenated AGNR edges,\cite{Wassmann} and consider also different structural isomers with respect to the two edges, as well as stereoisomers. Similar to the zigzag edges, the difference in total energy between isomers is less than 0.1~eV and we consider them to represent a single structure.

\begin{table}
 \caption{ \label{table:agnr7} The energetics for different armchair edge terminations: formation energy per unit length of the edge~($E_F$) and the binding energy per chlorine atom~($E_B$), both with radical chlorine~($_{\mathrm{Cl}}$) and with molecular chlorine~($_{\mathrm{Cl_2}}$) as the reference state.  Also the C-Cl bond length is reported. The ordering is based on edge stability with respect to molecular chlorine~($E_{F,\mathrm{Cl}_2}$). } 
\begin{tabular}{lccccc}
 \hline
 \hline
system & $E_{F,\mathrm{Cl}}$ &$E_{B,\mathrm{Cl}}$& $E_{F,\mathrm{Cl_2}}$ &$E_{B,\mathrm{Cl_2}}$&$d_{\mathrm{C-Cl}}$\\
  & \footnotesize{(eV/\AA{})} &\footnotesize{(eV/Cl)} &  \footnotesize{(eV\AA{})} &\footnotesize{(eV/Cl)} &\footnotesize{(\AA{})}\\
\hline
\footnotesize{HCl (cis)}	&-0.990 &-2.113 &-0.173 &-0.370 &1.83		\\
\footnotesize{HCl-H$_2$} 	&-0.574 &-2.450 &-0.165 &-0.706 &1.85		\\
\footnotesize{HCl (trans)}	&-0.978 &-2.088 &-0.162 &-0.345 &1.83 		\\
\footnotesize{H$_2$}		&-0.091 & -     &-0.091 & -     &  -  		\\
\footnotesize{HCl-Cl$_2$}	&-1.287 &-1.832 &-0.062 &-0.089	&1.80, 1.82	\\
\footnotesize{H$_2$-H$_2$-Cl$_2$-Cl$_2$}&-0.834&-1.781 &-0.017 &-0.037&1.78, 1.82,1.85 \\
\footnotesize{H$_2$-Cl$_2$}	&-0.806 &-1.721 &0.011  & 0.023 &1.86		\\
\footnotesize{H-Cl} 		&-0.342 &-1.463 &0.066  &0.281	&1.74		\\
\footnotesize{Cl-H-H-Cl}	&-0.327 &-1.396 &0.081  &0.348  &1.74		\\
\footnotesize{H-H-Cl-Cl}	&-0.320 &-1.368 &0.088  &0.375	&1.73		\\
\footnotesize{Cl}		&-0.694 &-1.483 &0.122  &0.260	&1.73		\\
\hline
\hline
\end{tabular}
\end{table}

\begin{figure*}
\includegraphics[width=0.55\textwidth, angle=90]{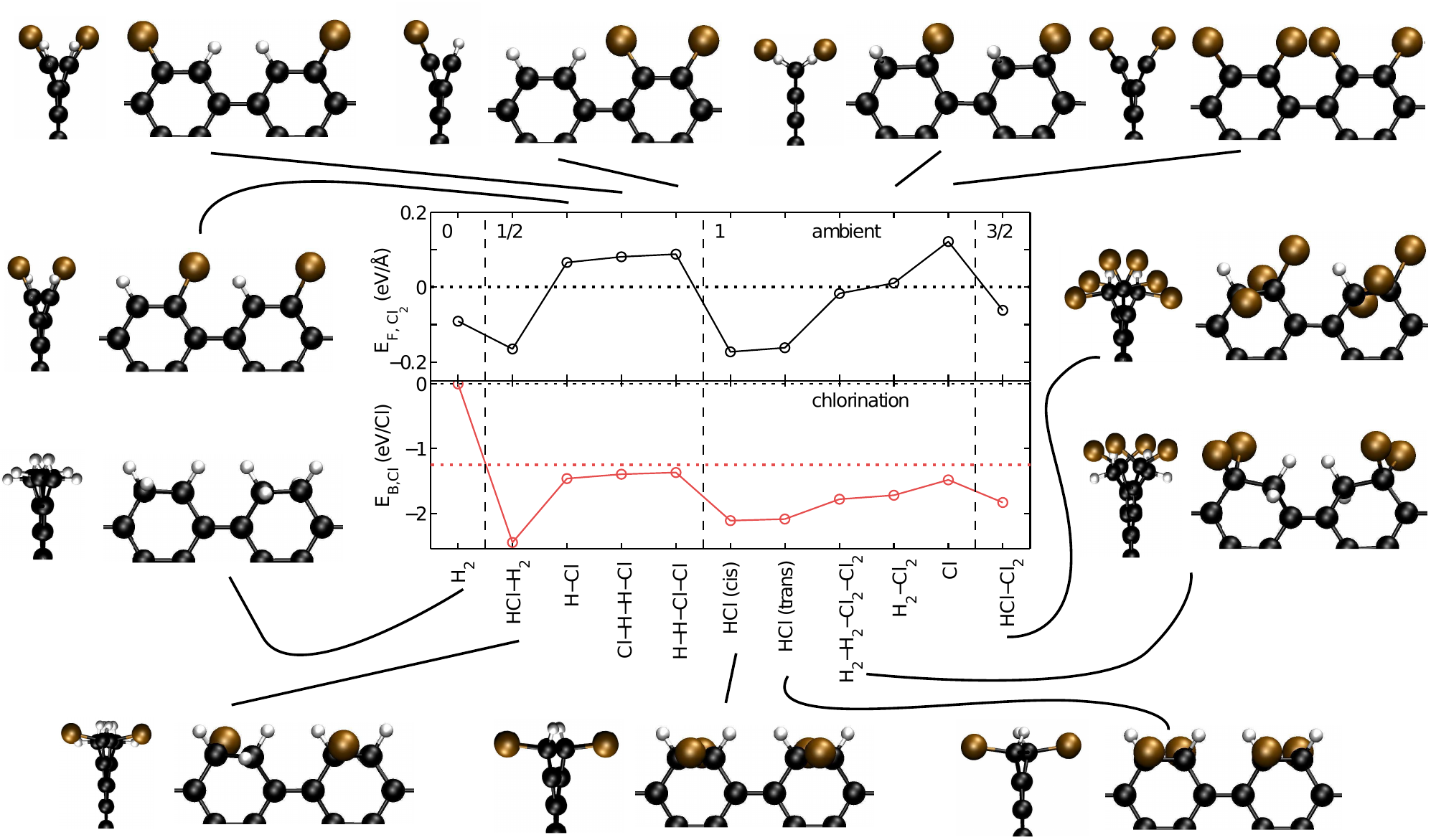}
\caption{\label{fig:agnr7} (Color online) The edge formation energy $E_{F, \mathrm{Cl}_2}$ in ambient conditions (upper panel) and binding energy per chlorine atom $E_{B, \mathrm{Cl}}$ (lower panel) in radical chlorine conditions for the different armchair terminations.In the upper panel, the dotted horizontal line shows stability with respect to pristine graphene ($E_{\mathrm{B,Cl_2}}=0$), and in the lower panel, the red dotted line shows the adsorption energy of a single chlorine atom on pristine graphene in the absence of substrate (12$\times$12 supercell, $E_{F, \mathrm{Cl}}=-1.25$~eV/Cl). The geometrical structures are connected to the graph from below (above) if they are (un)stable in ambient conditions. Vertical dashed lines separate groups of different degree of chlorination, expressed as chlorine atoms per edge carbon.  Atom colors as in Fig.~\ref{fig:zgnr8}.}
 \end{figure*}

The energetics of the lowest-energy chlorinated AGNR edges are given in Table~\ref{table:agnr7}, again in the order of energetic stability with respect to molecular chlorine ($E_{F, \mathrm{Cl}_2}$), and in Fig.~\ref{fig:agnr7} that also shows the geometries in the fashion of Fig.~\ref{fig:zgnr8}. Similar to zigzag edges, all chlorinated edges are stable in a radical chlorine environment in terms of edge formation energy $E_{F, \mathrm{Cl}}$, and in this environment, the chlorine binding energies are comparable for both edge types. In ambient conditions, however, only some of the AC geometries modifying the dihydrogenated edge are stable, and, in general, the AC edges are more stable than ZZ edges, having lower binding and formation energies. In addition, there are a larger number of stable configurations and as a matter of fact, even the doubly hydrogenated AC edge is more stable than the lowest-energy chlorinated ZZ edge. In ambient conditions, the edge formation energy of the favored AC edge is an order of magnitude lower than that of the favored ZZ termination,  $-0.173$~eV/\AA{} [HCl   (cis)] in comparison to $-0.016$~eV/\AA{} [H-Cl-HCl]. Actually, between these two formation energies there are four other AC edges that are energetically more stable than the lowest-energy zigzag edge. We do not find any spin polarization in chlorinated armchair structures. 

The degree of chlorination is higher for the most stable AGNR edge (1 Cl/edge carbon), compared to the most stable ZGNR configuration (2/3 Cl/edge carbon). Again, the low-energy structures modify the H$_2$ edge and have sp$^3$ hybridized carbon atoms with the C-Cl bond lengths 1.83-1.85~\AA{}.  The bonds between sp$^2$ carbons and chlorine are slightly stronger and thus shorter, the bond length being approximately 1.73~\AA{}. Notably, the binding energy per chlorine atom  in ambient conditions for the lowest-energy structures (Table~\ref{table:agnr7}, $E_{B,  \mathrm{Cl}_2}$) is not only lower than chlorine atom adsorption (0.499~eV/Cl) but also lower than the adsorption energy of molecular chlorine ($-0.125$~eV/Cl).

\subsection{Thermodynamic stability}

The DFT-calculated energies do not take into account the environment, for instance the pressure of the gaseous species and the finite temperature. It is intuitively clear that if there is only a very limited amount of a reagent  available, this may have an effect on the favored product. We may take these effects into account using \emph{ab initio} thermodynamics\cite{Wassmann, Seitsonen, Rogal} that allows us to compute the relative stabilities of different edge terminations as a function of the chemical potential of the gaseous species, the chlorine and the hydrogen. Although gaseous hydrogen is most likely scarcely available in the reaction conditions,\cite{Li}
the resulting phase diagram can be used as a qualitative rough guide to the effects of other forms of hydrogen, for instance hydrogen bound to the substrate. 

\begin{figure}
 \includegraphics[width = 0.45\columnwidth]{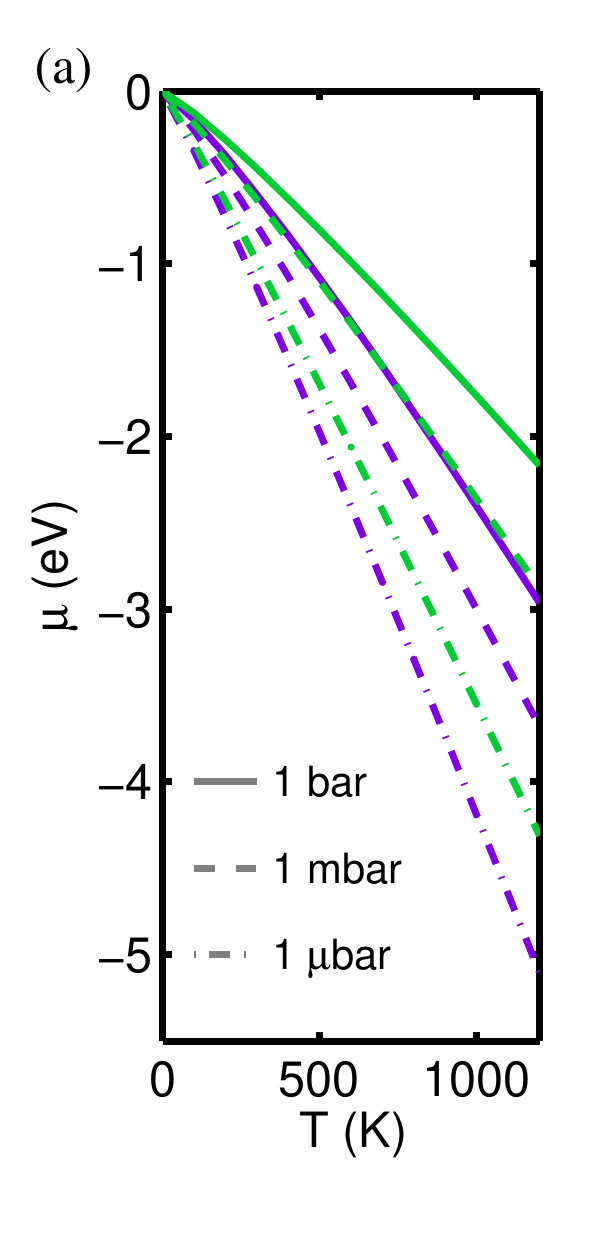}
 \includegraphics[width=0.45\columnwidth]{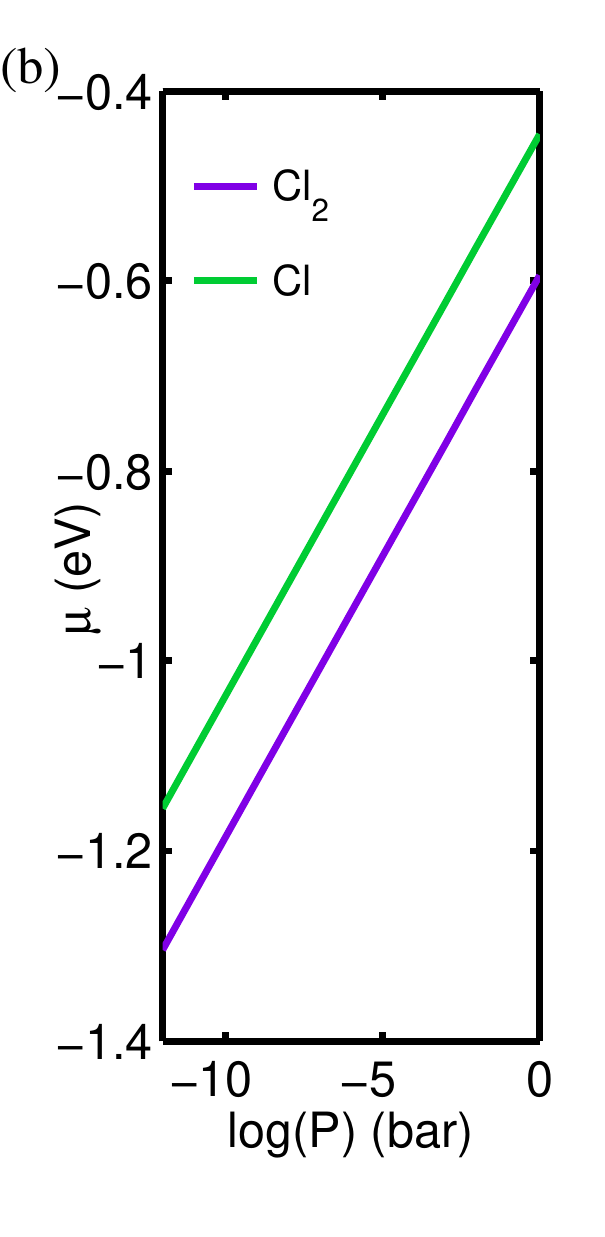}
\caption{\label{fig:chempot} (Color online) The chemical potential of molecular (violet/dark gray) and atomic (green/light gray) chlorine (a)~as a function of temperature at three values of $P$ (b)~as a function of partial pressure at room temperature ($T$=298~K). }
\end{figure}

The Gibbs energy of the edge formation is given by  
\begin{equation} \label{eq:Gibbs} \Delta G = E_{F,\mathrm{Cl}_2} -\frac{\rho_{\mathrm{H}}}{2} \mu_{{\mathrm{H}}_2}-\frac{\rho_{\mathrm{Cl}}}{2}\mu_{{\mathrm{Cl}}_2} \end{equation}
where the reference state for chlorine is chosen to be molecular chlorine, and $\rho_i$ gives the edge density of species $i$ and $\mu_i$ the chemical potential of species $i$. By using the formation energies calculated with respect to atomic chlorine and substituting $\rho_{\mathrm{Cl}}\mu_{{\mathrm{Cl}}_2}/2$ by $\rho_{\mathrm{Cl}}\mu_{\mathrm{Cl}}$, the Gibbs energy with respect to atomic chlorine can be determined. The chemical potential, given by
\begin{equation} \label{eq:mu} \mu_i (T) = H_i^{\circ} (T)-H_i^{\circ}(0)-TS_i^{\circ}(T)+k_BT\ln(P_i/P^{\circ}), \end{equation}
 where $H^{\circ}$ ($S^{\circ}$) is the enthalpy (entropy) in standard  pressure and $P_i$ the partial pressure of species $i$, can be can be  calculated using thermodynamical reference data.\cite{Chase,    Wassmann} Fig.~\ref{fig:chempot} shows the chemical potential of  gaseous atomic and molecular chlorine calculated using  Eq.~(\ref{eq:mu}) at  different temperatures and partial pressures. 

\begin{figure*}
 \includegraphics[width=0.24\textwidth]{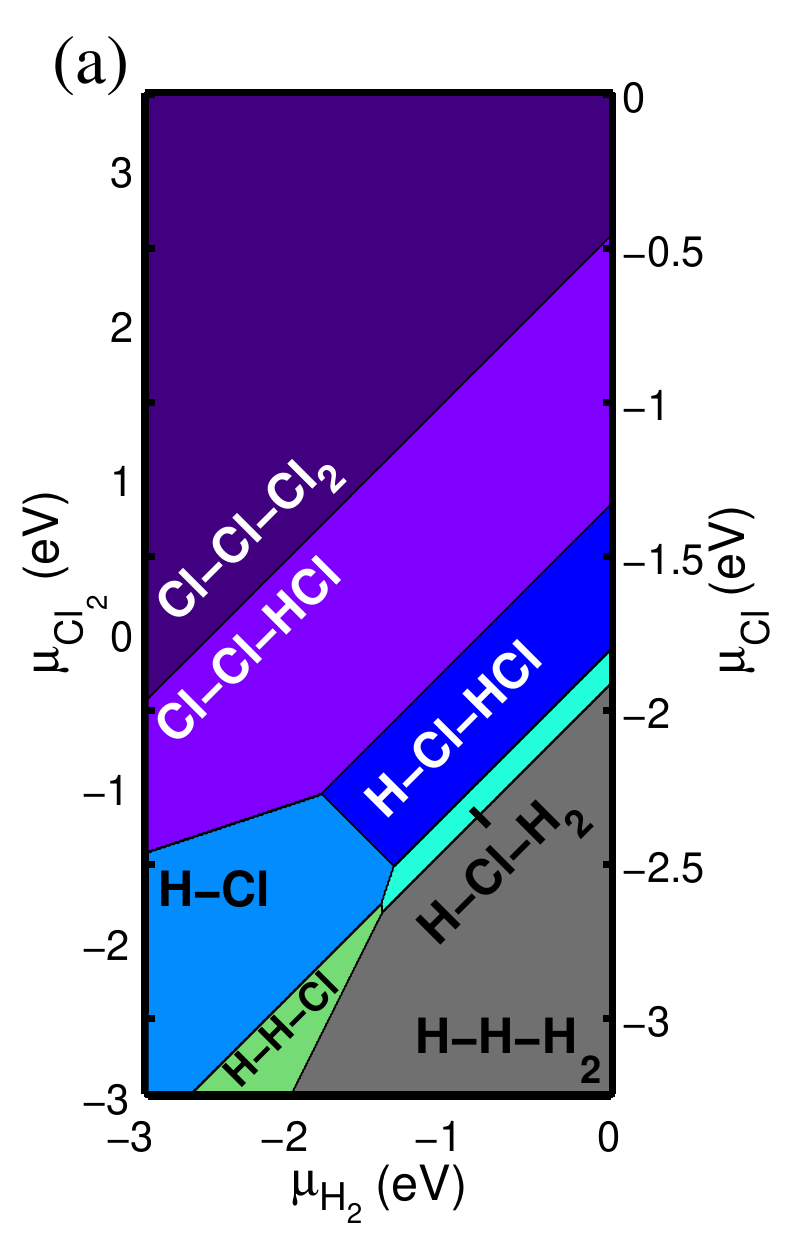}
\includegraphics[width=0.24\textwidth]{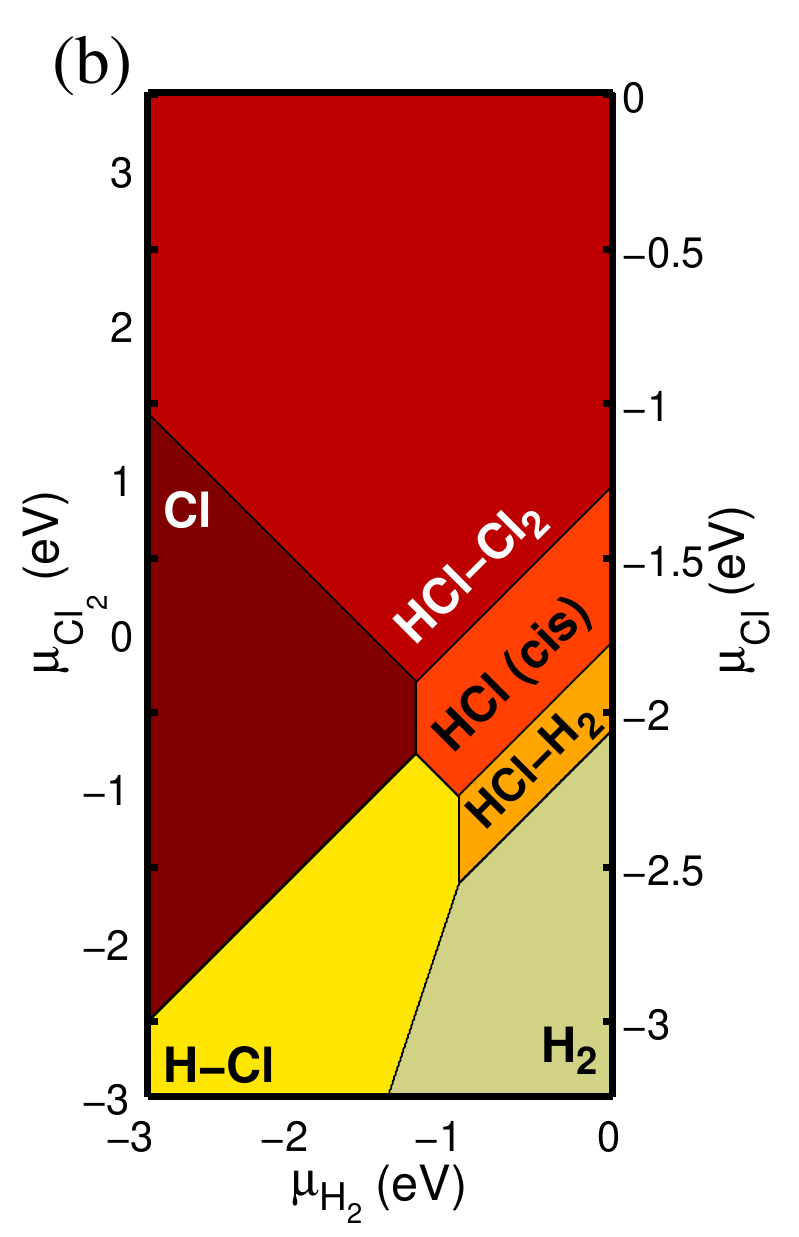}
\includegraphics[width=0.24\textwidth]{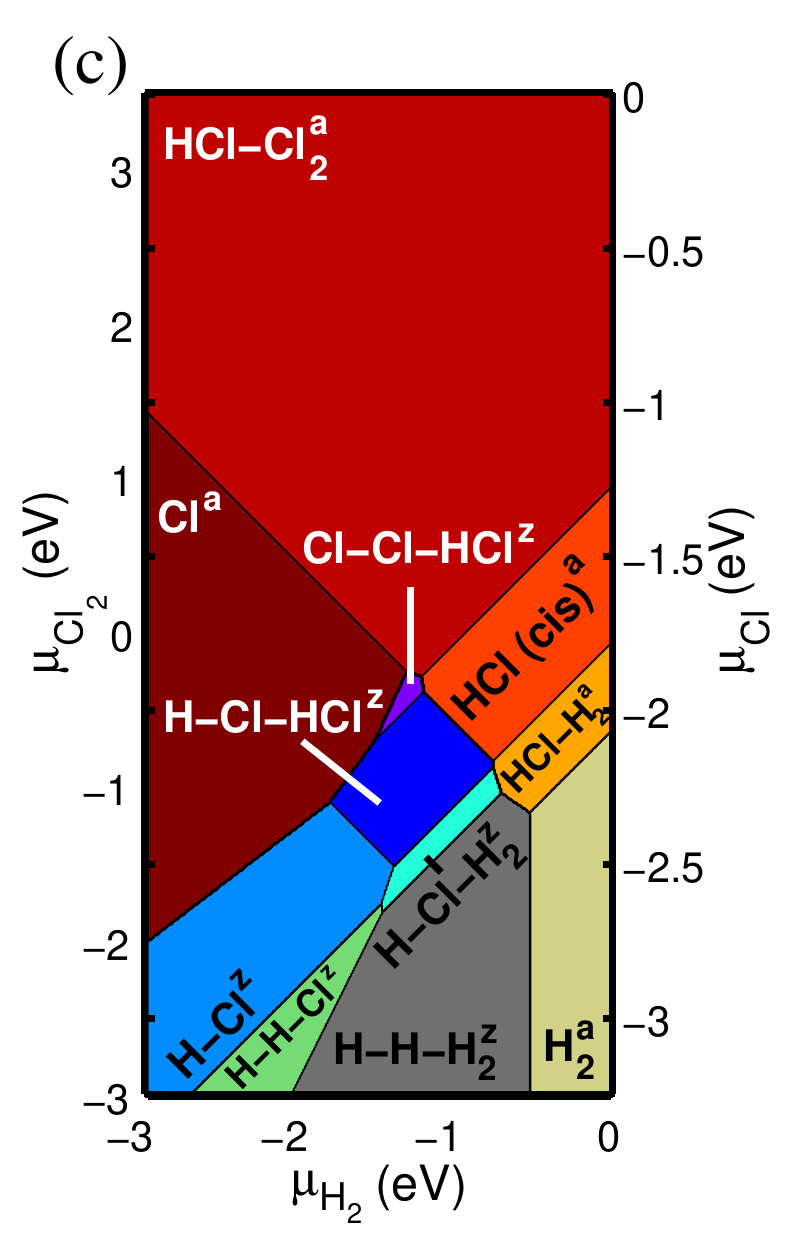}
\includegraphics[width=0.24\textwidth]{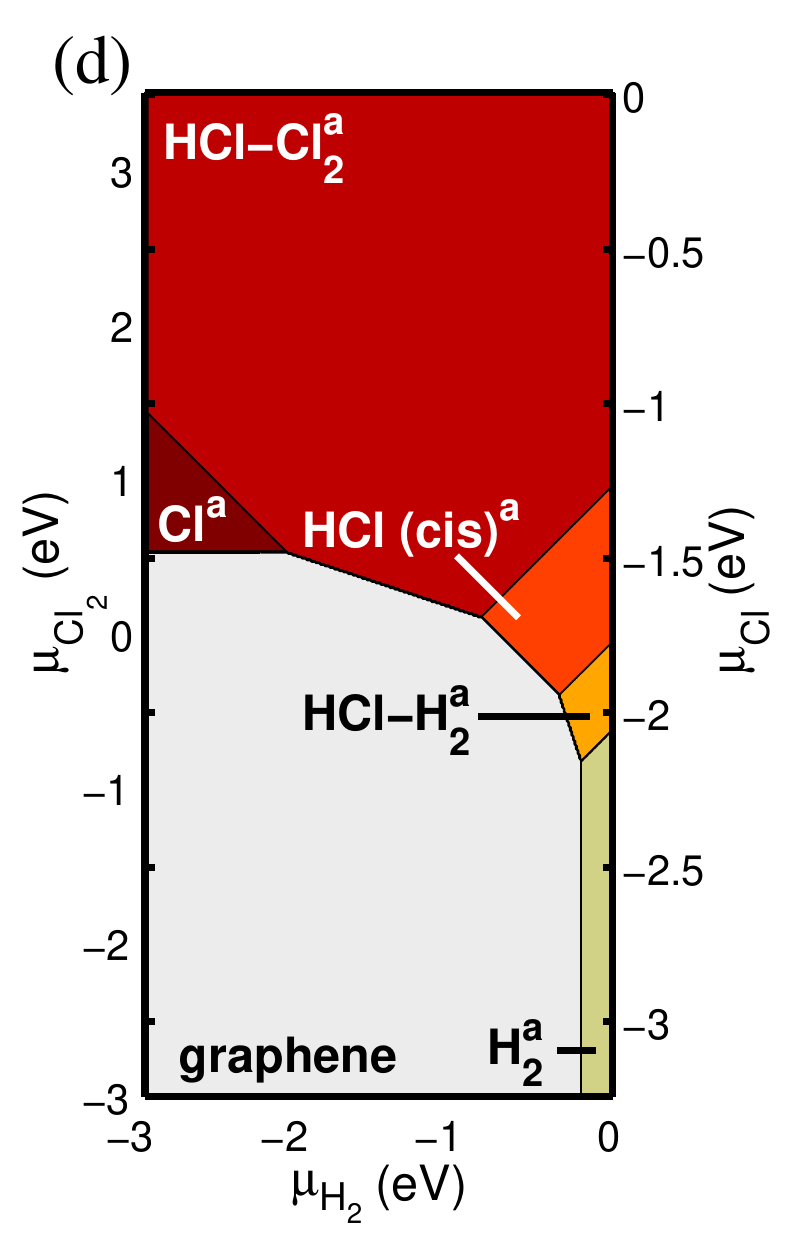}
 \caption{\label{fig:stability} (Color online) The phase stability diagrams showing the thermodynamically favored edge termination (with lowest $\Delta G$) in different hydrogen and chlorine environments, given by the chemical potential of the gaseous species Cl$_2$, Cl and H$_2$. (a)~Zigzag edges (b)~Armchair edges (c)~All edges  (d)~All edges including pristine graphene. }
\end{figure*}

As noted in the previous sections, in general the armchair-type terminations are more stable than their zigzag-type counterparts. Figs. \ref{fig:stability}(a) and \ref{fig:stability}(b) show separately the stability diagrams of the zigzag and armchair edges, that is the edge termination with the lowest $\Delta G$, as a function of the hydrogen and atomic chlorine chemical potentials. For comparison, also the chemical potential scale corresponding to molecular chlorine is given. As the chemical potential, and thus the partial pressure, of chlorine increases, the more densely chlorinated edge terminations become favored. Fig.~\ref{fig:stability}(c) combines subfigures (a) and (b). Zigzag edges are thermodynamically the most stable structures only at low chlorine and hydrogen chemical potentials.  
 
If $\Delta G$ of a termination is negative, an infinite graphene layer can be spontaneously broken into edges.\cite{Wassmann, Seitsonen} Thus, the possibility of pristine graphene being the favoured structure should be taken into account in the stability diagrams. Fig.~\ref{fig:stability}(d) shows a combined armchair-zigzag stability diagram including a region for pristine graphene. At low hydrogen and chlorine chemical potentials, that is at low pressures or concentrations, edge formation is not possible and pristine graphene is preferred. All chlorine- or hydrogen-containing terminations that are stable with respect to pristine graphene are of armchair type, and they are only favored over graphene at higher hydrogen and chlorine chemical potentials.  

Our approach does not consider chemical kinetics, or reaction barriers between different structures. In a real environment, metastable states are possible and this might allow some of the seemingly unstable chlorinated structures also to exist in ambient conditions. It is difficult to asses the experimental conditions used by Li~\emph{et al.}\cite{Li}, as they do not estimate their chlorine partial pressure. They, however, mention that the reaction is performed at room temperature, at 1~atm pressure in a mixture of radiated Cl$_2$ and N$_2$ so this gives an upper limit to the chlorine partial pressure.  As seen in Fig.~\ref{fig:chempot}(b), at room temperature this corresponds to a maximum chemical potential of approximately $-0.45$~eV that would give the armchair HCl-Cl$_2$ as the most stable edge configuration and allow its formation from pristine graphene.

\section{Discussion and conclusions}

We have studied the interaction of chlorine with graphene using density-functional theory. By comparing the energetics of different chlorine-containing edge terminations, both for armchair and zigzag graphene edges, to chlorine adsorption on pristine, freestanding graphene as well as graphene on the SiO$_2$ surface, we showed that chlorine binds to the edges rather than adsorbs onto the basal plane. In general, armchair edges are more readily chlorinated and the resulting structures are lower in energy compared with both atomic and molecular chlorine adsorption on pristine graphene. In addition, using \emph{ab initio} thermodynamics, we studied the effect of environment chlorine and hydrogen content on the stability of different edges and found that  pristine graphene may be spontaneously broken into edges in a sufficiently chlorine-rich environment. The presence of the silicon dioxide substrate lowers the adsorption energies onto the basal plane of graphene but even then the binding to the edges is energetically favored. 

In the experiment, Li~\emph{et al.}~\cite{Li} suggest that chlorine binds covalently to the basal plane of graphene, using the increase in the disorder-induced Raman D-peak as evidence, as well as XPS spectra showing covalent bonds between sp$^3$ hybridized carbon atoms and chlorine atoms.  We suggest that the nanodomains observed in the experiment by Li~\emph{et al.}\cite{Li} may actually be fractured graphene with chlorine-terminated edges. Our interpretation by no means contradicts {these experimental observations.} An enhanced reactivity upon chemical modification at graphene edges has previously been reported.\cite{Sharma} In addition, the Raman D-band is activated also through scattering from graphene edges.\cite{Casiraghi, Cancado, Cancado-Pimenta}  Carbon atoms that are sp$^3$-hybridized {and bind to chlorine}  are present in the lowest-energy edge terminations. As seen in our results  in the presence of a substrate, chlorine attachment onto the basal plane may be locally feasible if additional stabilization through the substrate is available. Local substrate-stabilized covalent binding could also explain the coverage, 8~atom-\%, estimated by Li~\emph{et al.} based on a XPS measurement that seems quite high to be achieved only through chlorinated edges.  

Li~\emph{et al.}~\cite{Li} also found that the average height of the chlorinated graphene increases from 0.9~nm to 1.1--1.7~nm upon the reaction with chlorine. As the carbon-chlorine bond length is only approximately 0.18-0.35~nm, the extremes corresponding to covalent bonds and physisorption,  binding to chlorine alone can not explain the increased height and roughness. It is possible that the formation of nanodomains is also related to surface morphology, so that graphene on higher regions or in regions with a certain surface termination react more readily, thus increasing the height variation.  Additionally, in fractured graphene, chlorine may be able to intercalate between the substrate and graphene, increasing the substrate-carbon distance. {Yet another possibility is that during the {edge formation}, the interaction with chlorine leads to upward bending of the edge regions, or partial overlap between adjacent edges.} Cleavage only in the high-symmetry armchair and zigzag directions is unlikely to occur in the reality,  and also bulky chlorinated hydrocarbon groups [of type (CH$_{\mathrm{x}}$Cl$_{\mathrm{2-x}}$)$_n$CH$_{\mathrm{y}}$Cl$_{\mathrm{3-y}}$] could form at the edges.  Even if the Raman maps of the chlorine-reacted graphene do not show spatial variation, the resolution of the measurement with a beam of 1~$\mu$m in diameter is much larger than the lateral dimension of the observed domains, 30--50~nm.\cite{Li} 

Our results show that alternative explanations than the simple basal plane adsorption, such as a combination of chlorine-containing graphene edges and substrate effects studied in this work are needed to explain graphene chlorination in experiments. Additionally, defects are likely to be important as reactive sites in the basal plane but this is a topic of future work. 

\acknowledgments

We acknowledge computational resources from CSC-IT Center for Science Ltd. and useful discussions with Andreas Uppstu. M.I. acknowledges the financial support from the Finnish Doctoral Programme in Computational Sciences FICS.

\bibliography{Cl_graphene_final}

\begin{thebibliography}{46}
\bibitem [{rep()}]{repulsion}%
  \BibitemOpen
  \href@noop {} {}\bibinfo {note} {Attraction and formation of Cl$_2$ molecules
  is forbidden through the periodic boundaries.}\BibitemShut {Stop}%
\end{thebibliography}

\end{document}